# Mid-infrared optical coherence tomography with MHz axial line rate for real-time non-destructive testing

Satoko Yagi[1,*], Takuma Nakamura[2,*], Kazuki Hashimoto[2], Shotaro Kawano[1], and Takuro Ideguchi[1,2,**]

[1]Departmnet of Physics, The University of Tokyo, Tokyo 113-0033, Japan

[2]Institute for Photon Science and Technology, The University of Tokyo, Tokyo 113-0033, Japan

[*]These authors contributed equally

[**]ideguchi@ipst.s.u-tokyo.ac.jp

**Abstract**

Non-destructive testing (NDT) is crucial for ensuring product quality and safety across various industries. Conventional methods such as ultrasonic, terahertz, and X-ray imaging have limitations in terms of probe-contact requirement, depth resolution, or radiation risks. Optical coherence tomography (OCT) is a promising alternative to solve these limitations, but it suffers from strong scattering, limiting its penetration depth. Recently, OCT in the mid-infrared (MIR) spectral region has attracted attention with a significantly lower scattering rate than in the near-infrared region. However, the highest reported A-scan rate of MIR-OCT has been 3 kHz, which requires long data acquisition time to take an image, unsatisfying industrial demands for real-time diagnosis. Here, we present a high-speed MIR-OCT system operating in the 3-4 μm region that employs the swept-source OCT technique based on time-stretch infrared spectroscopy. By integrating a broadband femtosecond MIR pulsed laser operating at a repetition rate of 50 MHz, we achieved an A-scan rate of 1 MHz with an axial resolution of 11.6 μm and a sensitivity of 55 dB. As a proof-of-concept demonstration, we imaged the surface of substrates covered by highly scattering paint coatings. The demonstrated A-scan rate surpasses previous state-of-the-art by more than two orders of magnitude, paving the way for real-time NDT of industrial products, cultural assets, and structures.

**Introduction**

Non-destructive testing (NDT) techniques have evolved to meet the increasing demands for product quality and safety in various industrial sectors[1]. These methods detect hidden defects within materials without causing damage, making them highly effective for inspecting large-scale manufacturing lines, cultural assets, buildings, and structures. Various imaging techniques are available for NDT, including those based on ultrasonic[2], terahertz[3], and X-ray[4] modalities. However, each has its limitations. Ultrasonic inspection requires probe contact and offers limited depth resolution, ranging from a few hundred micrometers to several millimeters. Terahertz imaging eliminates the need for probe contact but suffers from low lateral and depth resolution due to the long wavelength. Meanwhile, X-ray inspection techniques provide a high depth resolution of several micrometers, but they require bulky detection equipment and carry the risk of radiation exposure.

In recent years, optical coherence tomography (OCT) in the mid-infrared (MIR) spectral region has emerged as a promising technology to address the limitations of current NDT methods[5–14]. OCT measures the interference of broadband light using a Michelson interferometer, enabling non-contact and non-destructive determination of internal reflections within materials[15]. While conventional OCT systems employ near-infrared light and are primarily used for medical diagnoses, especially cross-sectional retinal imaging, they can also be adapted for NDT purposes[16]. However, light scattering significantly limits their penetration depths to micrometer scale. Since scattering cross section has a significant wavelength dependence (e.g., $\lambda^{-4}$ for Rayleigh scattering, where $\lambda$ denotes wavelength), using longer wavelengths can vastly enhance penetration depth. Specifically, MIR-OCT in the 3-4 μm spectral band has attracted attention for non-contact NDT because this wavelength region is in the atmospheric window where carbon dioxide and water vapor have low absorption cross sections. In past research, MIR-OCT has showcased high-resolution and deep imaging capabilities in scattering media, including ship coatings[12], paintings[9], papers[13], credit cards[7], and ceramics[5,7,9,14].

However, the current state-of-the-art measurement speed of MIR-OCT does not satisfy the requirements for industrial applications. Prior studies on high-speed MIR-OCT utilized the spectral-domain OCT (SD-OCT) method, incorporating a diffraction grating and a line sensor[6,9,10]. The highest recorded A-scan rate was 3 kHz, which was limited by the frame rate of the line sensor[10]. This A-scan rate leads to a measurement time of at least several minutes for taking an image of 1000×1000 pixels on the lateral plane. Hence, substantial enhancement is required for real-time measurements, such as in-line inspections of industrial products and in-situ evaluations during processing. State-of-the-art laser-scanning nonlinear microscopes utilize high-speed beam scanning systems composed of galvanometric and resonant scanners for real-time video-rate measurements. Considering using the high-speed scanning system for MIR-OCT, the pixel dwell time can be reduced to 0.1-1 μs to take an image of 1000×1000 pixels on the lateral plane, which corresponds to an A-scan rate of 1-10 MHz, indicating considerable potential for improvement.

In this study, we introduce a high-speed MIR-OCT system that leverages the swept-source OCT (SS-OCT) method, incorporating the recently developed time-stretch infrared spectroscopy (TSIR)[17,18]. As the spectrum acquisition rate of TSIR aligns with the pulse repetition rate, it allows for a significantly increased A-scan rate. To achieve both high axial resolution and high-speed TSIR-OCT measurements, we developed a broadband femtosecond MIR pulsed laser with a repetition rate of 50 MHz. The spectral bandwidth of the laser, which determines the axial resolution of OCT measurements, is 1000 nm, spanning from 3200 to 4200 nm in the atmospheric window. Using the developed laser, we demonstrated broadband TSIR-OCT that offers a high axial resolution of 11.6 μm and a record A-scan rate of 1 MHz with a sensitivity of 55 dB. For a proof-of-concept demonstration, we imaged the surface of substrates painted by highly scattering media, which conventional OCT systems in the near-infrared region struggle to measure.

**Results**

We developed a MIR laser suitable for TSIR-OCT, which requires a high repetition rate and a broad spectrum in the 3-4 μm region. To meet these criteria, we chose difference frequency generation (DFG) with a femtosecond mode-locked Yb fiber laser as a master oscillator. Figure 1(a) depicts a schematic of the developed laser. Our amplified Yb-fiber laser generates 230-fs pulses centered at 1037 nm at a repetition rate of 50 MHz (YLMO, Menlo Systems)[19]. The pulses from the oscillator are split into two branches. One branch is used for pump pulses in the DFG process, and the other is for signal pulses. The signal pulses are generated through spectral broadening by supercontinuum (SC) generation in a 14-cm-long photonic crystal fiber (PCF) (IXF-SUP-5-125-1050-PM, iXblue). The fiber length is optimized to broaden the spectral bandwidth while suppressing the temporal broadening of the pulses. The generated SC pulses, spanning from 650 nm to 1560 nm, are spectrally filtered with a longpass filter with a cut-on wavelength of 1300 nm (FELH1300, Thorlabs), which is displayed as an inset in Fig. 1(a). The signal pulses are collimated with a lens to have a beam diameter of 2 mm. Using a dichroic mirror, the signal pulses are spatially combined with the pump pulses, whose beam diameter is 1.6 mm. Then, these pulses are focused with an f=19 mm achromatic lens onto a 0.5-mm-long MgO-doped periodically poled lithium niobate (MgO:PPLN) crystal with a poling period of 29.8 μm (FOPMIR-MA-C-0.5, HC Photonics). The average power of the signal and pump pulses are 120 and 700 mW, respectively. The polarization and path-length differences between the signal and pump pulses are adjusted using a half-wave plate and a delay line, respectively. The DFG process generates a broadband MIR spectrum spanning from 3200 to 4200 nm, which is limited by the phase-matching condition of the PPLN crystal. The average power of the generated MIR pulses is around 10 mW. Note that the generated spectral bandwidth of 1000 nm is much broader than previous demonstrations with similar settings, which had about 170 nm[20], 500 nm[21], or 240 nm[22] but with higher average powers. In the previous studies, the spectral bandwidths were limited by those of Raman solitons generated in PCFs with shorter zero-dispersion wavelengths (ZDWs). In contrast, we used a PCF with a longer ZDW of 1050 nm, closer to the pump wavelength, leading to a smooth supercontinuum spectrum and a simple temporal waveform[23], which allows for utilizing the broader spectrum determined by the phase-matching condition.

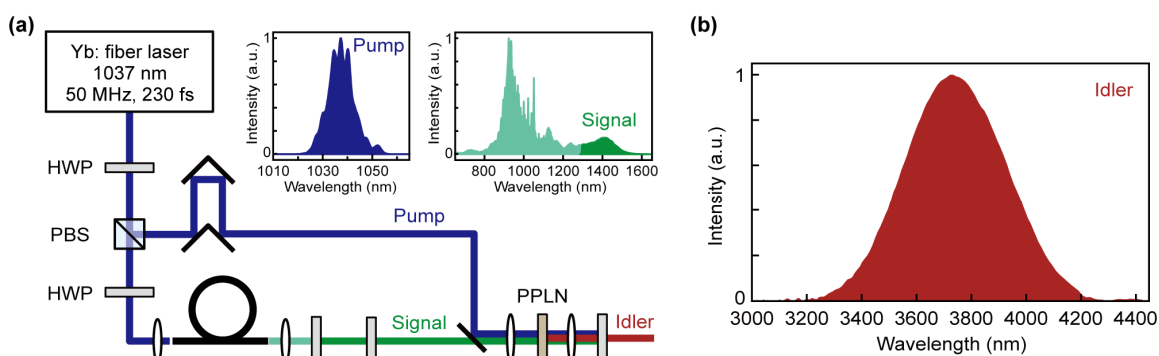

**Fig. 1 Broadband MIR laser for TSIR-OCT.** (a) A schematic diagram of the DFG source based on a Yb: fiber laser. Insets show the spectra of (left) pump pulses and (right) signal pulses, respectively. The former is the same as that of the oscillator. The latter represents a portion of the SC spectrum, highlighted in dark green in the figure. The whole SC spectrum is shown in light green. (b) The MIR idler spectrum generated via the DFG process. HWP: half-wave plate, PBS: polarizing beamsplitter, PCF: photonic crystal fiber, LPF: longpass filter, DM: dichroic mirror, PPLN: periodically poled lithium niobate.

Figure 2 illustrates a schematic diagram of our TSIR-OCT system and its underlying principle. The system comprises a Michelson interferometer and a time-stretch spectrometer (Fig. 2(a)). The MIR pulses from our developed laser are directed into a Michelson interferometer and split into two with a CaF$_2$ wedged beamsplitter (BSW510, Thorlabs). In the sample arm, the pulses are focused onto a sample with an f=50 mm BaF$_2$ lens, which has a flatter dispersion profile than CaF$_2$ in the 3-4 μm region. The spot size in the focal plane is 106 × 122 μm$^2$, which is evaluated by knife-edge measurements. The MIR pulses are spatially scanned by a 2-D raster scanner consisting of a 12-kHz resonant scanner (Cambridge Technology) and 120-Hz galvanometric scanner (Cambridge Technology) to obtain 3-D images. The field of view (FOV) and the number of pixels in the lateral direction are 3.6 × 2.0 mm$^2$ and 40 × 22 pixels, respectively. The FOV is optimized so that the pixel size (90 × 90 μm$^2$) closely matches the beam spot. In the reference arm, the MIR pulses are focused onto a mirror using a BaF$_2$ lens. The MIR pulses from the two arms are spatially recombined at the beamsplitter and directed to the TSIR spectrometer. We inserted a 5-mm CaF$_2$ window into the sample arm for dispersion compensation of the beamsplitter.

As a spectrometer, we employed a recently developed upconversion TSIR (UC-TSIR) technique[18]. In UC-TSIR, the MIR pulse spectra are converted to near-infrared (NIR) spectra through the DFG process so that time-stretch detection can be operated in the telecom region with a low-loss optical fiber and a high-speed photodetector. For wavelength conversion, we employed the DFG process. In the previous study, we used a continuous-wave laser as a pump source for DFG to ensure a one-to-one spectral transfer, which required a long nonlinear crystal to gain conversion efficiency, limiting the spectral bandwidth[18]. In this study, to achieve an efficient conversion with a thin nonlinear crystal for broadband DFG while keeping a one-to-one spectral transfer, we utilized ps pulses as a pump source for DFG, which were generated by spectrally filtering a portion of pulses from the master oscillator. The ps pulses centered at 1037.5 nm are amplified to 490 mW with a homemade two-stage Yb-doped fiber amplifier[24], and the spectral width is adjusted to 0.44 nm, determining the spectral resolution of the upconversion process. An f=20 mm CaF$_2$ lens focuses the spatially combined MIR and pump pulses onto a 0.5-mm-thick MgO: PPLN crystal with a poling period of 29.8 μm (FOPMIR-MA-C-0.5, HC Photonics). The DFG converts the MIR spectrum to the NIR region, centered around 1450 nm, with a bandwidth of approximately 150 nm and an average power of 30 μW. Then, the MIR and ps pulses are filtered off using a long-pass filter with a cut-on wavelength of 1200 nm.

These upconverted NIR pulses are coupled into a 10-km single-mode fiber (SMF-28 Ultra, Corning) with a coupling efficiency of 60% and temporally stretched to around 15 ns. The total insertion loss of the fiber is 3.5 dB. The stretched pulses are detected with a 10-GHz InGaAs photodetector (RXM10AF, Thorlabs) and amplified with two 14-dB, 20-GHz RF amplifiers (HMC460LC5, ANALOG DEVICES). The detected signals are digitized using a 16-GHz oscilloscope (WaveMaster816Zi-B, Teledyne Lecroy) at a sampling rate of 40 GSamples/s. The depth of field (DOF) of OCT measurements is determined by the spectral resolution of the system. In UC-TSIR-OCT, the spectral resolution can be determined by either the total detection bandwidth or the linewidth of the picosecond laser for upconversion. In our system, the former (10 GHz) corresponds to the DOF of 2.3 mm, and the latter corresponds to the DOF of 1.2 mm. The wavelength axis of the measured TSIR waveform is calibrated by a reference spectrum measured with an optical spectrum analyzer at a spectral resolution of 0.2 nm. For MIR-OCT measurements, we averaged TSIR waveforms 50 times to reduce the intensity noise of the laser, resulting in an A-scan rate of 1 MHz.

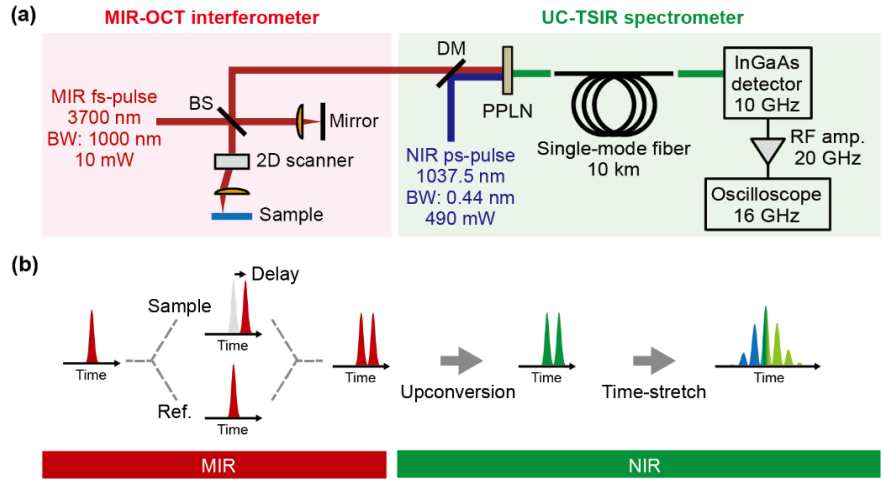

**Fig. 2 Schematic diagram and principle of TSIR-OCT.** (a) A schematic representation of TSIR-OCT. The system comprises an MIR-OCT interferometer and a UC-TSIR spectrometer. (b) Principle of TSIR-OCT in the time domain. For illustrative simplicity, the figure only displays a single reflected pulse from the sample. The delayed pulses from both the sample and reference arms generate the spectral interferometric fringes, which are measured by the time-stretch spectrometer in the time domain. BS: beamsplitter, DM: dichroic mirror, PPLN: periodically poled lithium niobate, BW: bandwidth.

To validate the one-to-one spectral conversion through the upconversion DFG process, we compare an input MIR-OCT spectrum with an output upconverted NIR spectrum. We placed a mirror at the sample position and observed a spectral fringe pattern produced by the double pulses from the sample and reference arms. Figures 3(a) and 3(b) display an MIR spectrum taken with a homemade Fourier-transform infrared (FTIR) spectrometer and an upconverted NIR spectrum measured with an optical spectrum analyzer, respectively. The comparison indicates that the upconversion process achieves a one-to-one spectral conversion from the MIR to the NIR region. The shorter wavelength end of the NIR spectrum is slightly diminished, which we attribute to the conversion bandwidth being constrained by the phase-matching condition in the PPLN crystal.

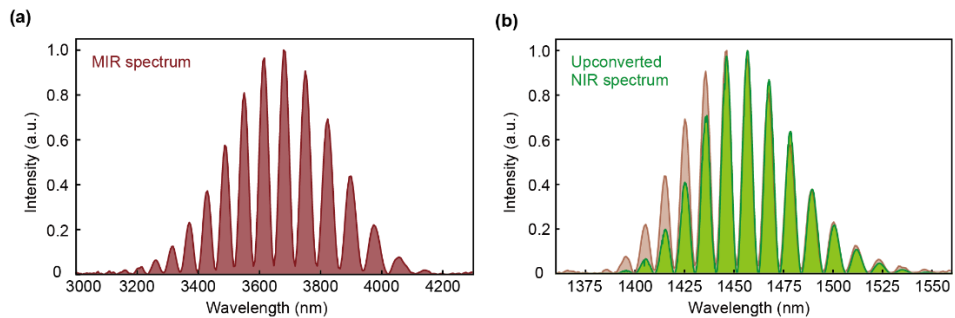

**Fig. 3 Validation of the one-to-one spectral conversion through the upconversion DFG process.** (a) an MIR spectrum, and (b) an upconverted spectrum of the fringe pattern obtained by producing double pulses from the sample and reference arms. The red-shaded plot shown in (b) is an expected spectrum when the MIR spectrum shown in (a) is converted by an ideal DFG process. The spectral fringes align well, confirming the upconversion process successfully achieves a one-to-one spectral conversion. A minor attenuation on the shorter wavelength end of the NIR spectrum results from the restricted conversion bandwidth, determined by the phase-matching condition in the PPLN crystal.

Next, we assess the specifications of our TSIR-OCT system. Figure 4(a) provides an illustration of the test measurement. We employ a gold mirror as a test sample and adjust its depth position using a mechanical stage. Figure 4(b) presents OCT fringe patterns captured at various depth positions. Each waveform is averaged over 50, corresponding to an A-scan rate of 1 MHz. The figure displays fringe patterns with various periods, confirming its capacity to convey depth information. Figure 4(c) shows the Fourier-transformed spectra of the captured temporal waveforms, revealing distinct narrow lines at different frequencies based on depth positions. The horizontal and vertical axes represent the depth position of the mirror and sensitivity, respectively. The sensitivity is calculated using the standard methodology for OCT, outlined in previous studies[25]. The sensitivity is evaluated as 55 dB for a signal at a depth of 100 μm. The noise was evaluated by taking the standard deviation of the noise floor at a corresponding depth of 400-800 μm. A roll-off declination is observed at larger depths. The origin of the roll-off is attributed to the reduction of visibility, mainly due to the beam divergence, which can be mitigated by using a focusing lens with a longer focal length. Figure 4(d) illustrates the linewidth of the signal in the frequency domain, which is evaluated as 11.6 μm, determining the depth resolution of the system. It is worth noting that this demonstration also showcases the broadest UC-TSIR spectroscopy, extending to 1000 nm (744 cm$^{-1}$) in the 3-4 μm region, which is 37 times larger bandwidth than the previous study (20 cm$^{-1}$)[18]. This suggests its potential applicability for high-speed, broadband molecular spectroscopy as well.

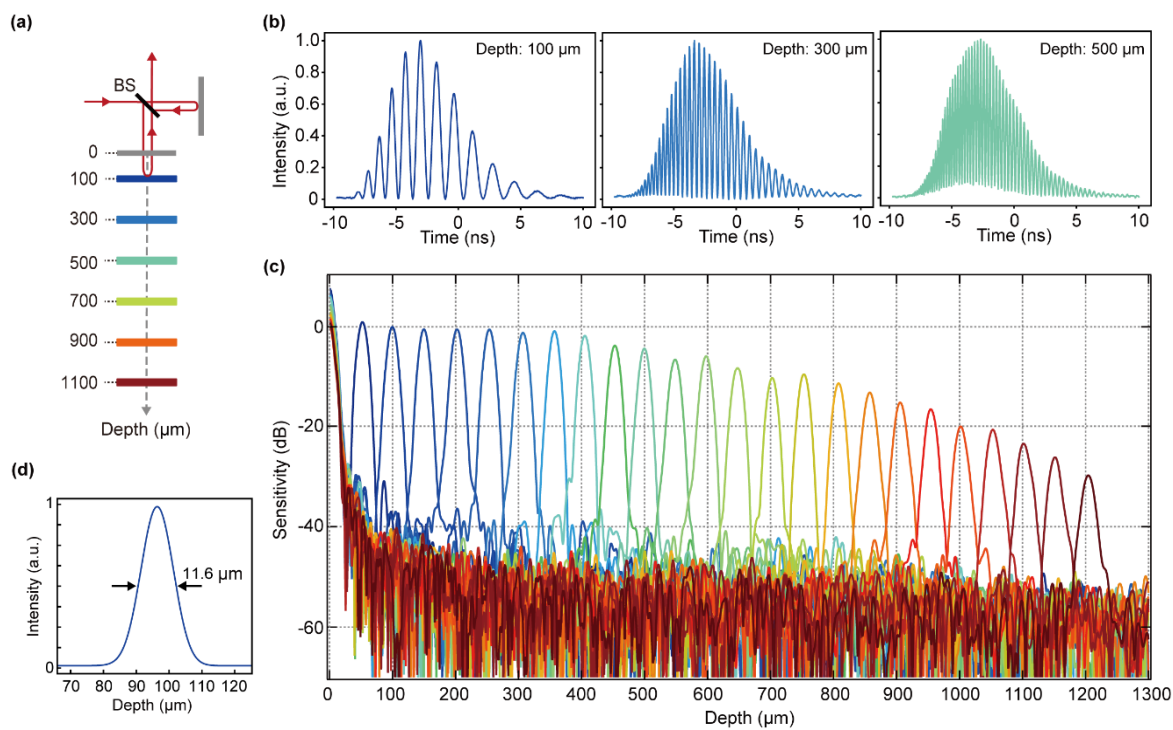

**Fig. 4 Performance evaluation of the TSIR-OCT system.** (a) A schematic illustration of the mirror positions in the sample arm. (b) Measured time-stretched waveforms corresponding to the mirror depths of 100, 300, and 500 μm. (c) Fourier-transformed spectra of the time-stretched waveforms. The horizontal and vertical axes represent depth and sensitivity, respectively. (d) A line profile of a Fourier-transformed spectrum shown in (c). The vertical axis uses a linear scale, showing a full width at half maximum of 11.6 μm. BS: beamsplitter.

Finally, we applied our system to NDT of paint coatings on metal substrates as a proof-of-concept demonstration for measuring inside highly scattering media. We prepared two painted samples: aluminum (Al) substrates sprayed with either red or white colored materials used for automobile painting. These paints contain small particles designed to scatter or absorb visible light. Figures 5(a) and (b) present the measured 3D OCT images of the samples with microscopy images. The images show the edge of the coatings, where the surfaces of the Al substrate and the paint coatings are observed. Figures 5(c) and (d) display OCT cross-sectional images captured by our TSIR-OCT and a commercial NIR-OCT system operating in a 900 nm wavelength region (GAN621, Thorlabs), respectively. Figures 5(e) and (f) exhibit the A-scan plots. Our TSIR-OCT captures the surfaces of the substrates beneath the highly scattering coatings. The thicknesses of the coatings are estimated to be 89 and 149 µm, respectively, derived from half the measured optical path difference (OPD) divided by the paint medium's refractive index (1.5). In contrast, the results obtained with the commercial NIR-OCT system display prominent noise due to the intense scattering by the coatings, rendering them unable to discern the Al surfaces beneath. Figure 6 showcases a 3D OCT image of another painted sample with a scratch on the substrate surface. The image clearly reveals the scratch beneath the scattering coating, which is otherwise invisible over the coating. The FOV spans 3.6 mm × 2.0 mm in the lateral direction. The recording time for the 3D images (40×22×172 pixels) was 1 ms. Our demonstration validates the rapid measurement capability of our TSIR-OCT, enabled by its unprecedentedly high A-scan rate.

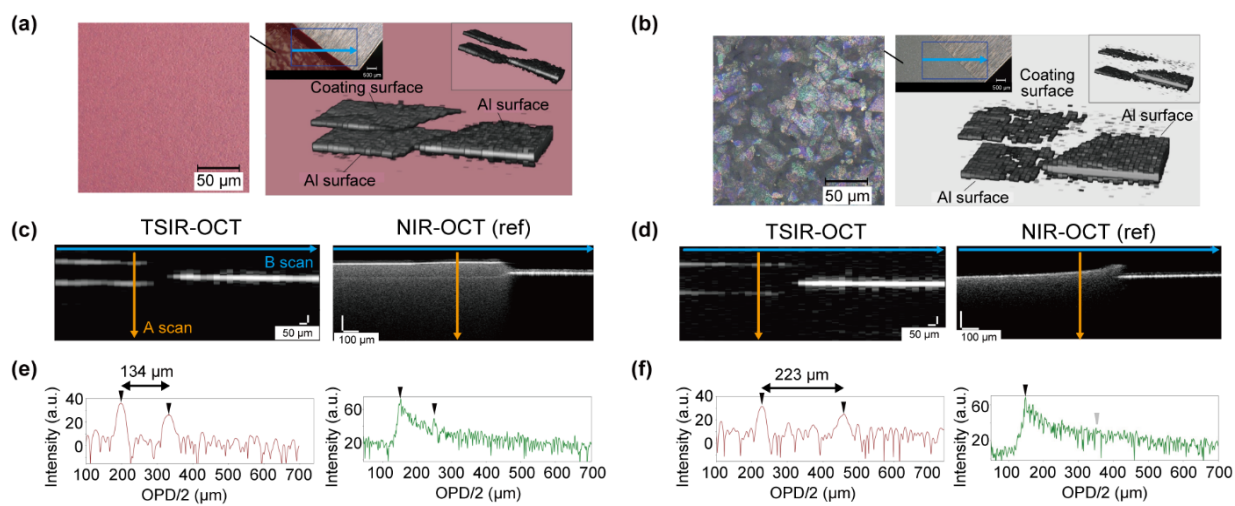

**Fig. 5 NDT of highly scattering paint coatings on aluminum substrates.** (a)(b) 3D image of a painted aluminum substrate captured with the TSIR-OCT system. The inset presents a microscope view. (c)(d) B-scan cross-sectional depth images taken at the position indicated by the blue arrow in (a). The left and right panels represent results from TSIR-OCT and NIR-OCT, respectively. (e)(f) A-scan cross-sectional images captured at the location marked by the orange arrow in (c) and (d). The left and right panels represent measurements by TSIR-OCT and NIR-OCT, respectively. OPD: optical path difference.

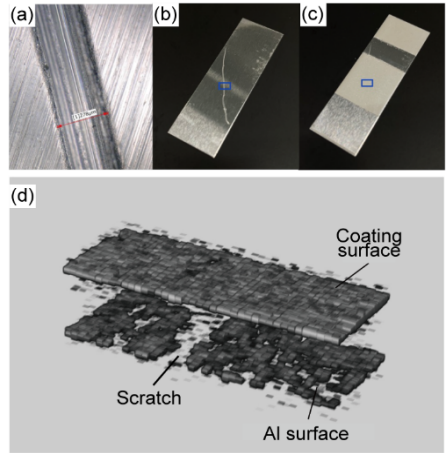

**Fig.6 NDT of paint coating on a scratched aluminum substrate.** (a) Magnified microscope image of a scratch. (b) Microscope view of the scratched aluminum substrate. (c) Microscope image of the substrate coated with white paint. (d) 3D image captured using the TSIR-OCT system.

**Discussion**

It is valuable to compare the performance of our TSIR-OCT to previously demonstrated MIR-OCT systems and discuss potential improvements. Our current A-scan rate of 1 MHz is already 333 times higher than the prior art of 3 kHz[10]. It can be further increased to 50 MHz by fully utilizing the repetition rate of our MIR laser system. However, in the current setup, the DFG laser exhibits significant pulse-to-pulse intensity noise, which necessitates 50 times averaging. We could solve this issue by using a reported technique to suppress the intensity noise by finely adjusting the delay between the input pulses for DFG[26]. Our present axial resolution of 11.6 μm is slightly lower than the highest resolution of 8.6 μm demonstrated in the previous study[10]. The axial resolution can be enhanced by broadening the spectral bandwidth of the DFG laser, either by using a shorter or a chirped PPLN crystal. The DOF of the current system is 1.2 mm, which is limited by the linewidth of the picosecond pulses for upconversion. This can be improved by filtering the spectrum of the picosecond pulses to narrow it down. The sensitivity in our demonstration is 55 dB for a 1 μs measurement time. In contrast, the cutting-edge method achieved 65 dB[10] for 0.33 ms, or 82 dB for 10 ms[6]. Considering the averaging effect on sensitivity, TSIR-OCT could deliver 80 dB for 0.33 ms or 95 dB for 10 ms, which outperforms earlier demonstrations. Our sensitivity is presently limited by detector noise, which can be overcome by capturing more photons within the linearity range of the detector. It is achievable by enhancing the average power of the MIR pulses or by improving the efficiency of the upconversion process.

It is worth estimating the feasibility of imaging for thicker samples. We consider alumina ceramics and zirconia ceramics as examples, where the former is used in electronics as an insulator, while the latter is employed for artificial teeth. We calculated the scattering loss of these materials in both the MIR (3500 nm) and NIR (1000 nm) regions using an open-source program named BHMIE, which accounts for both Mie and Rayleigh scatterings. The parameters needed for this calculation are sourced from the literature[5]. The assumed pore volumes and diameters for alumina and zirconia ceramics are (1%, 0.4 μm), and (0.2%, 0.2 μm), respectively. The penetration depths, determined by a 1/e decrease in light intensity, are 2 mm in the MIR (3500 nm) and 1.2 μm in the NIR (1000 nm) for alumina ceramics.

For zirconia ceramics, the penetration depths are 43 mm in the MIR and 230 μm in the NIR. These results suggest the potential capabilities of TSIR-OCT for NDT of mm- to cm-scale highly scattering objects. It is worth noting that the DOF of the system, which is determined by the spectral resolution, must be enlarged by roughly ten times for cm-scale depth measurements. In our TSIR-OCT, it is readily possible to improve the spectral resolution with a 10-fold narrower linewidth of the picosecond pulses for upconversion and 10-fold longer optical fiber for time-stretching, along with a 10-fold lower repetition rate. A similar high-resolution TSIR measurement was demonstrated in our previous study[18].

## Conclusion

We have demonstrated an MIR-OCT system with a record-high A-scan rate of 1 MHz, utilizing a homemade broadband MIR laser source and the TSIR spectroscopy technique. Our MIR source produces a broadband spectrum that covers an atmospheric window from 3200 to 4200 nm, ensuring a high axial resolution of 11.6 μm. The UC-TSIR provides a one-to-one conversion of the MIR-OCT spectra to the NIR region, facilitating high-speed SS-OCT measurements with low-loss time-stretching and efficient detection in the telecom region. Our proof-of-concept demonstration on measuring painted substrate showed its superiority in OCT imaging of highly scattering media compared to traditional NIR OCT systems. The achieved A-scan rate of 1 MHz paves the way for video-rate 3D OCT imaging through real-time recording and processing of digitized waveforms.

## Funding

JSPS KAKENHI (20H00125, 21K20500, 23H00273), Research Foundation for Opto-Science and Technology, Nakatani Foundation, UTEC-UTokyo FSI Research Grant Program.

## Acknowledgments

We thank Hiroyuki Shimada for supporting the experiment and Makoto Shoshin for commenting on the manuscript.

## Author contributions

T.I. conceived the concept of the work. S.Y. and T.N. designed the system. S.Y. and T.N. constructed the optical systems with the help of K.H.. S.Y. performed the experiments and analyzed the data with the help of T.N.. S.Y. and S.K. took data with the NIR-OCT system. T.I. supervised the work. T.I. wrote the manuscript with inputs from all the authors.

## Competing interests

T.N., K.H., and T.I. are the inventors of the filed patent related to the UC-TSIR technique.

## Data availability

The data provided in the manuscript are available from the corresponding author upon reasonable request.